\documentclass[aps,prl,showpacs,twocolumn,groupedaddress,
amsfonts,amssymb, superscriptaddress]{revtex4}

\oddsidemargin

\usepackage{graphicx}%
\usepackage{dcolumn}

\begin{document}

\title{Critical Casimir force in $^4$He films: confirmation of finite-size scaling}

\author{A.~Ganshin}

\author{S.~Scheidemantel}
\affiliation {Department of Physics, The Pennsylvania State
University, University Park, Pennsylvania, 16802}

\author{R.~Garcia}
\email{garcia@wpi.edu} \affiliation{Department of Physics, Worcester
Polytechnic Institute, Worcester, MA 01609}

\author{M.~H.~W.~Chan}
\affiliation {Department of Physics, The Pennsylvania State
University, University Park, Pennsylvania, 16802}

\date{\today}

\begin{abstract}
We present new capacitance measurements of critical Casimir
force-induced thinning of $^4$He films near the superfluid/normal
transition, focused on the region below $T_{\lambda}$ where the
effect is the greatest. $^4$He films of 238, 285, and 340 \AA\
thickness are adsorbed on N-doped silicon substrates with roughness
$\approx 8 {\AA}$. The Casimir force scaling function $\vartheta $,
deduced from the thinning of these three films, collapses onto a
single universal curve, attaining a minimum $\vartheta = -1.30 \pm
0.03$ at $x=td^{1/\nu}=-9.7\pm 0.8 {\AA}^{1/\nu}$. The collapse
confirms the finite-size scaling origin of the dip in the film
thickness. Separately, we also confirm the presence down to $2.13 K$
of the Goldstone/surface fluctuation force, which makes the
superfluid film $\sim 2 {\AA}$ thinner than the normal film.
\end{abstract}



\maketitle

An important focus in condensed matter physics is understanding how
the properties of a thermodynamic system evolve as its size is
shrunk to ever smaller dimensions. Near a continuous phase
transition or critical point, the theory of finite-size scaling
offers a testable prediction. According to finite-size scaling, the
correction to the free energy per unit area of a planar film of
thickness $d$ due to confinement of critical fluctuations has a
simple, universal form \cite{fdg}
\begin{equation}
 \delta F_{12} = \frac{k_B T_c} {d^2} \Theta_{12}
 ( d/\xi) \label{eq:one}.
\end{equation}
where $T_c$ is the transition temperature and the correlation length
$\xi = \xi_{0}|t|^{-\nu}$ measures the spatial extent of
fluctuations in the bulk. $t = T/T_c-1 $ is the reduced temperature.
The scaling function $\Theta_{12}$ is predicted to be a
dimensionless, universal function of the ratio $d/\xi$ and the
boundary conditions that the order parameter satisfies at the
confining interfaces.

While finite-size scaling is applicable to all critical systems, the
most rigorous experimental tests to date have focused on the scaling
behavior of the specific heat anomaly of $^4$He near the superfluid
transition \cite{lipa,gasp}. This is due to the nearly-ideal,
impurity-free nature of liquid $^4$He and the low-sensitivity of
this system to gravitational rounding errors. For the superfluid
transition, $T_c = T_{\lambda} = 2.1768 K$ and $\nu= 0.67016  \pm
0.00008$ \cite{Goldner}. For a $57 \mu m$ thick film, the magnitude
and temperature dependence of the specific heat is found to be in
reasonable agreement with finite-size scaling predictions
\cite{lipa}. However, for films $500-7000 {\it \AA}$ thick, the
situation is not as clear-cut. The temperature-dependence of the
specific heat is as expected from the universal $d/\xi$ dependence
in Eq.~(\ref{eq:one}). The maximum specific heat occurs at a common
value $x=(d\xi_0/\xi)^{1/\nu}=td^{1/\nu}=-9 \pm 1 {\AA}^{1/\nu}$ for
all films, where the negative $x$ refers to the maximum occurring
\textit{below} $T_{\lambda}$. However, the \textit{magnitude} of the
specific heat shows an unexpected, systematic non-collapse
\cite{gasp}.

The critical Casimir force is another fundamental manifestation of
finite-size scaling that is open to experimental testing. Just as
the Casimir force between two conducting plates arises due to the
confinement of zero-point electromagnetic fluctuations between the
plates \cite{cas}, a completely analogous \textit{thermodynamic}
Casimir force is expected between the substrate and vapor interfaces
of adsorbed liquid films, due to the confinement of critical
fluctuations within the thickness of the film
\cite{fdg,castheory,crcasimir,tribinary,williams}. The
theoretically-predicted critical Casimir force per unit area
\begin{equation}
 f = - \frac {\partial \delta F_{12}}{\partial d} =
 \frac {k_B T_c} {d^3} \vartheta_{12} ( d/\xi ) \label{eq:two}
\end{equation}
where Casimir scaling function $\vartheta(z) = 2\Theta(z) - z
\partial \Theta/ \partial z$. Because in $^4$He films the superfluid
order parameter vanishes at both film interfaces, the critical
Casimir force is attractive ($\vartheta < 0$) \cite{castheory},
producing a dip in the equilibrium film thickness near
$T_{\lambda}$. The existence of this dip, first observed by
\cite{hallock}, has been confirmed in a quantitative experiment
using as substrates five pairs of capacitor plates made of polished
Cu set at different heights above bulk liquid helium
\cite{crcasimir}. The interpretation of this experiment is
complicated by the roughness of the Cu surface, which changes the
effective areas of the Cu plates and makes it impossible to
accurately determine the film thickness. AFM scans over $2500 \mu
m^2$ areas of the surfaces show they are not ideal, with 10-130 \AA\
rms roughness and occasional micron-deep scratches and dust
particles. $\vartheta$ is calculated using DLP theory \cite{cole}
and by assuming that the Cu surfaces at different heights are flat.
The result of the experimental analysis \cite{crcasimir} is that the
scaling function $\vartheta$ exhibits a behavior suggestively
similar to the specific heat. The minimum in the dip occurs at a
common value $x=-9.2 \pm 0.2 {\AA}^{1/\nu}$ for all the films
257-423 \AA\ thick. The temperature-dependence is exactly that
expected from from the $d/\xi$ dependence in Eq.~(\ref{eq:two}), but
the magnitude of $\vartheta $ shows an unexpected non-collapse, the
minimum of $\vartheta$ increasing systematically from -1.85 to -1.4
as $d$ increases from 257 \AA\ to 423 \AA\ \cite{crcasimir}. To
address whether this systematic trend in the magnitude of
$\vartheta$ is an artifact due to the non-ideal surface or is truly
related to the non-collapse observed for the specific heat, we have
undertaken improved capacitance measurements of the critical Casimir
force similar to \cite{crcasimir} but using flat N-doped silicon
surfaces with roughness $\approx$ 8 \AA.

\begin{figure}[th]
\vskip 0.2 cm \includegraphics[width=70mm]{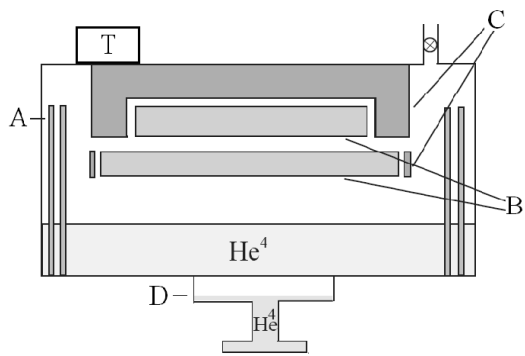} \vskip 0.2
cm \noindent \caption{The sketch of the experimental cell. The level
of the bulk $^4$He at the bottom of the cell is measured using an
annular capacitor (A). The $^4$He film is adsorbed on the two
silicon plates (B) attached to electrically-grounded copper guard
rings (C) held 0.2 $mm$ apart by Cu shim spacers. The temperature of
the cell was measured using a germanium thermometer $T$ attached to
the top of the cell, which is calibrated \textit{vs.} the $^3$He
vapor curve and the lambda fixed point device (D) at the bottom.
\label{fig1} }
\end{figure}

A sketch of the experimental cell machined from oxygen-free high
conductivity Cu is presented in Fig. 1. Two silicon (100) wafers
highly-doped with phosphorous (1-5 $m\Omega/ \square $ ) are
configured as parallel plates forming a capacitor with a gap $G =
235 \mu m$. According to a tapping-mode AFM, the rms roughness of
the wafers is 8 \AA. To minimize surface roughness, dust particles,
and scratches, the experimental cell is washed, dried, assembled and
sealed in the Penn Sate Nanofabrication facility, a class $10$ clean
room. Virgin wafers, completely intact, are rubber-cemented into the
Cu guard rings used to position the electrodes. The top electrode is
a 1 inch wafer, and the bottom is 2 inch.  To minimize error from
the fringe field, the top and bottom guard rings are grounded and
the 1 inch wafer is placed at virtual ground in the AC bridge
circuit used to measure the capacitance $C$ \cite{crcasimir,pestak}.
To determine $T_{\lambda}$, a fixed point device anchored to the
cell bottom is used, following the procedure described by
\cite{maid}.

The temperature control scheme of our experiment is similar to that
of the original experiment of \cite{crcasimir}. A needle valve is
used to close the helium fill line just above the cell. The data are
taken with the cell slowly drifting through the lambda point, at
$10-40 \mu K/h$ near $T_{\lambda}$ where equilibration takes longer
and at $70-300 \mu K/h$ below $T_{\lambda}$. We use two thermal
control stages. The first outer stage is maintained at constant
temperature with less than $50 \mu K$ noise. To achieve a uniform
temperature drift rate, we apply heat to a second stage, just above
the cell. After dosing helium into the cell, we typically observe
signs of capillary condensation, where liquid droplets condense in
the gap between the silicon electrodes. To get rid of these
droplets, we very slowly ($100- 300 \mu K/h$) thermally cycle the
cell through $T_{\lambda}$, each time looking for a distinctive drop
in capacitance that signals the flowing of liquid from the gap. This
procedure is repeated until a reproducible $C(T)$ dependence is
obtained.

To calculate the film thickness $d$ from the measured $C(T)$, we
model $C(T)$ as the equivalent capacitance due to three dielectric
layers added in series: adsorbed film, vapor phase, and adsorbed
film, obtaining
\begin{equation}
d=\frac{G}{2}\left( \frac{1}{\epsilon _{\it vapor}} - \frac
{1}{\epsilon (T)}\right)/\left( \frac {1}{\epsilon _{\it vapor}} -
\frac {1}{\epsilon _{\it film}}\right)\label{eq:three}
\end{equation}
where, if $C_0(T) $ is the temperature-dependent empty capacitance,
the effective dielectric constant $ \epsilon (T) =C(T)/C_0(T)$. As
in \cite{crcasimir}, the dielectric constant of the film
$\epsilon_{\it film} = 1.05760\pm 0.00005$ and the dielectric
constant of the vapor $\epsilon _{\it vapor}$ is calculated using
the Clausius-Mossotti equation, taking the molar polarizability of
helium to be $0.123296 \pm 0.000030 cm^3/{\it mol}$. The vapor
density is calculated from the pressure $P(T)$, using the second
virial coefficient $B(T)$ from \cite{virial}.

\begin{figure}[th]
\includegraphics[width=80mm]{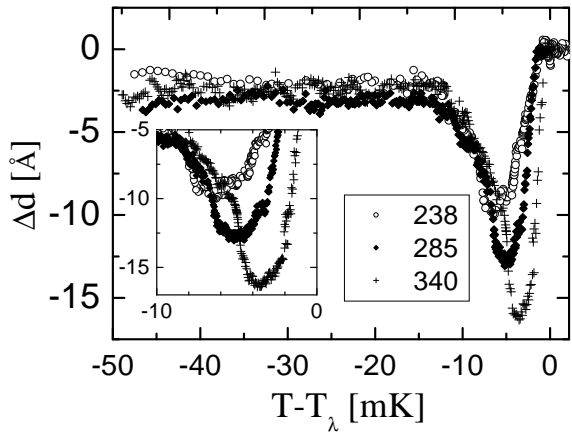}
\vskip 0.1 cm \noindent \caption{The thinning of the film plotted
vs. temperature $T$: (a) the data shown over a wide range of
temperature, (b) a blow up near the minimum. The films are labeled
by the thickness in \AA\ in the region above $T_{\lambda}$ where the
Casimir force is negligible. \label{fig2} }
\end{figure}

The temperature-dependence of $C_0$ is due to a small linear
increase in $G$ caused by a combination of liquid surface tension
acting on the Cu spacers and differential thermal contractions among
the various materials that make up the capacitor, including between
the silicon wafer and the rubber cement underneath. In our data
analysis, we assume that $C_0(T)=C_0(T_{\lambda})(1 - 3.5 \times
10^{-5}(T-T_{\lambda}))$. This results in a temperature-independent
$d$ for all films for $T$ sufficiently \textit{ above as well as
below} $T_{\lambda}$. Each time we dose liquid into the cell to make
a new film, we characteristically observe an additional small shift
in $G$ (and $C_0$) on the order of $50 ppm$. To correct for this, we
adjust $C_0(T_{\lambda})$ in order to obtain the
theoretically-predicted thickness in the regime above $T_{\lambda}$
where the critical Casimir force is negligible and the equilibrium
thickness $d$ on the silicon wafer is expected to be determined
solely by a competition between temperature-independent van der
Waals and gravitational forces. In this regime, the film thickness
is given by \cite{cole}
\begin{equation}mgh =  \frac{\gamma_0}{d^3}
\left(1+\frac{d}{d_{1/2}} \right)^{-1} \label{eq:four}
\end{equation}
where, on the left side $mgh$ is the chemical potential due to
gravity, fixed by the height $h$ above the bulk liquid, where $g$ is
the gravitational acceleration and $m$ the atomic weight of helium.
On the right side is a simplified expression for the chemical
potential due to van der Waals forces, where $\gamma_0 \approx 1950
K {\it \AA}^3$ and $d_{1/2} \approx 230 {\it \AA}$ are
substrate-specific interpolation parameters that characterize the
net attraction of the helium to the silicon, including retardation
effects \cite{cole}. The parameters $\gamma_0$ and $d_{1/2}$ are
approximate, ignoring the effect of the small 20 \AA\ natural oxide
layer on the silicon. Nevertheless, the error is estimated to be
less than 5\% or 10 \AA\ and the same for all the films studied.

\begin{figure}[th]
\includegraphics[width=80 mm]{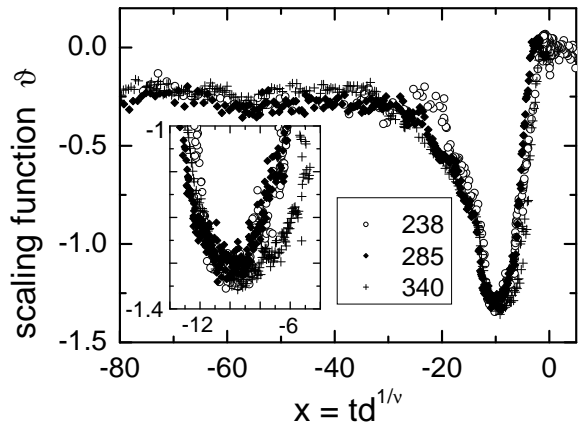}
\vskip 0.1 cm \noindent \caption{The scaling function $\vartheta $
vs the scaling variable $x$. The minimum occurs at $x=-9.7 \pm 0.8
{\AA}^{1/\nu}$, for all three films. \label{fig3} }
\end{figure}

In Fig.~\ref{fig2}, we show the measured change in the film
thickness in response to the temperature-dependent Casimir force
near and below $T_{\lambda}$, for three different values of the
height $h=$ 15.00, 8.01, and 4.22 $\pm 0.05 mm$. The films are
labeled by their thicknesses above $T_\lambda$ calculated from
Eq.~(\ref{eq:four}); namely 238, 285, and 340 $\pm 10 {\AA}$. As
seen previously \cite{crcasimir}, due to the Casimir force, thicker
films exhibit larger dips which occur closer to $T_{\lambda}$.
Including the additional contribution to the chemical potential from
the critical Casimir force \cite{castheory}, the equilibrium film
thickness is expected to be given by
\begin{equation}mgh =  \frac{\gamma_0}{d^3}
\left(1+\frac{d}{d_{1/2}} \right)^{-1}+ \frac{k_{B}
T_{\lambda}V}{d^3}\vartheta(d/\xi) \label{eq:five}
\end{equation}
where $V = 45.81 {\AA}^3/atom $ is the specific volume of liquid
$^4$He, and $\vartheta$ is the dimensionless scaling function for
the Casimir force. The observed dip in $d$ is due to $\vartheta <0$,
i.e. an attractive Casimir force between the substrate and vapor
interfaces, as expected due to the superfluid order parameter
satisfying Dirichlet boundary conditions at the two film interfaces
\cite{castheory}.

In Fig.~\ref{fig3}, we show the Casimir force scaling function
calculated using Eq.~(\ref{eq:five}) and the data of
Fig.~\ref{fig2}. Because it was necessary to disrupt data collection
every 2.5 days to transfer cryogen, each curve, which takes about 2
weeks to complete, actually consists of 4-5 overlapping data sets
that are spliced together; this results in additional noise and a
small discrepancy very close to $T_{\lambda}$ not present in the
earlier work \cite{crcasimir}. Nevertheless, within the scatter, all
three data sets collapse onto a single curve, with a minimum value
of $\vartheta = -1.30 \pm 0.03$ at $x=-9.7 \pm 0.8 {\AA}^{1/\nu}$.
The collapse of the data verifies that the dip in the film thickness
near $T_{\lambda}$ is due to fluctuation-induced forces
\cite{castheory}. It is noteworthy that the measured $\vartheta$
shows quantitative agreement with the $\vartheta$ obtained
previously \cite{crcasimir} for $423 {\AA}$ thick $^4$He films on
Cu, but disagrees with the results obtained for thinner films that,
presumably, would be more sensitive to surface non-idealities. These
results suggest the non-collapse is the result of inadequate
corrections for the effects of surface roughness and not due to
$\vartheta$ depending on the additional off-coexistence variable
$hd^{\Delta/\nu}$ where $\Delta/\nu = 2.47$ \cite{castheory}. This
is expected to have important implications for the analysis of
specific heat and wetting experiments \cite{gasp,lipa,balibar}.

The new measurements, which focus on obtaining data near the minimum
of the dip and over a wide range below $T_{\lambda}$, confirm an
additional important aspect of earlier experiments: for all the
films studied, we find the superfluid film is $ \sim 2{\AA}$ thinner
than the normal film down to 2.13K. Experiments indicate that the
onset of superfluidity in the films occurs somewhere between $x=-7$
and $ -12 {\AA}^{1/\nu}$ \cite{gasp,qcm}. Thus it has been suggested
that the thinner superfluid film is caused by Casimir forces due to
fluctuations involving superfluidity in the film, such as Goldstone
modes, second and third sound \cite{kardar}. As seen from Fig. 3,
the thinning in the superfluid film is consistent with an
asymptotic, low-temperature value of the Casimir force $\approx
-(0.30 \pm 0.10)k_B T/d^3$. This force is marginally larger than the
$-0.15 k_B T/d^3$ force predicted by \cite{kardar}.

In summary, the current experiment confirms the validity of
finite-size scaling formula for the critical Casimir force in
adsorbed $^4$He films between 230 and 340 \AA\ thick. Measurements
down to $\sim 2.13 K$ also show the presence of an additional,
non-critical, attractive fluctuation-induced force in the superfluid
film. Our study underscores the importance of smooth surfaces for
these types of measurements. For future work, it would be desirable
to test the scaling of Casimir forces in a much wider thickness
range that overlaps the range covered by specific heat measurements.
We would like to thank A. Maciolek, R. Zandi, J. Rudnick, M. Krech,
D. Dantchev, S. Dietrich, M. Kardar, S. Balibar, G. Williams, F. M.
Gasparini, and M. W. Cole for useful discussions and comments.
This research was supported by NASA grant NAG3-2891.

\end{document}